**ORIGINAL ARTICLE**

# Fault-Tolerant Backup Clustering Algorithm for Smart Connected Underwater Sensor Networks

Abdulazeez Femi Salami [1], Emmanuel Adewale Adedokun [2], Habeeb Bello-Salau [2] and Bashir Olaniyi Sadiq[2]

[1] Department of Computer Engineering, University of Ilorin, Nigeria.
[2] Department of Computer Engineering, Ahmadu Bello University, Zaria.

**ABSTRACT** – This paper addresses poor cluster formation and frequent Cluster Head (CH) failure issues of underwater sensor networks by proposing an energy-efficient hierarchical topology-aware clustering routing (EEHTAC) protocol. In this paper, fault-tolerant backup clustering (FTBC) algorithms and multi-parameter cluster formation (MPCF) model were developed for the EEHTAC operation. The MPCF model tackles the issue of poor cluster formation performance by integrating multiple parameters to achieve effective clustering process. The FTBC algorithms tackle the issue of frequent CH failures to avoid interruption in data transmission. Performance of the MPCF model was evaluated using normal, high-fault, and high routing overhead network scenarios. Performance metrics employed for this analysis are temporal topology variation ratio (TTVR), CH load distribution (CLD), and cluster stability (STB). Obtained results show that operating with a CH retention period of 90s achieves better CH duty cycling per round and improves the MPCF process with values of 25.69%, 55.56%, and 60% for TTVR, CLD, and STB respectively. Performance of the FTBC-based EEHTAC was evaluated relative to Energy-balanced Unequal Layering Clustering (EULC) protocol. Performance indicators adopted for this evaluation are routing overhead ($\Omega$), end to end delay ($\Delta$), CH failures recovered (CFR), CH failures detected (CFD), received packets ($\theta$), and energy consumption ($\Sigma$). With reference to the best obtained values, EEHTAC demonstrated performance improvement of 58.40%, 29.94%, 81.33%, 28.02%, 86.65%, and 54.35% over EULC variants in terms of $\Omega$, $\Delta$, CFR, CFD, $\theta$, and $\Sigma$ respectively. Obtained results displayed that the MPCF model is efficient for cluster formation performance and the FTBC-based EEHTAC protocol can perform effectively well against an existing CBR protocol.

## INTRODUCTION

Recent innovations in Internet of underwater things (IoUT) have enhanced the performance of underwater sensor networks (UWSNs) [1, 2, 19, 31]. IoUT can be conceptually viewed as an extension and application of IoT to aquatic environments through a specialized UWSN [3, 20, 32]. Emerging UWSNs are expected to perform collaborative target monitoring in order to fully realize a smart connected network of subaquatic sensors with intelligent computing, massive data processing, self-learning and adaptive decision-making capabilities [4, 5, 23, 26]. Due to these technological enhancements, IoUT solutions are practically considered as an indispensable ingredient and essential asset for realizing smart cities [6, 7, 36, 47, 51].

However, an obstacle the full realization of smart connected underwater sensor network (SC-UWSN) is the harsh underwater environment [33, 35, 37, 41, 43, 48]. In order to address this challenge, different types of routing protocols have been proposed by researchers; out of which cluster-based routing (CBR) protocols have demonstrated more versatility with respect to large-scale UWSN applications [7, 8, 34, 45, 49]. In CBR, the deployed nodes are classified into cluster members (CMs) and cluster heads (CHs) based on one or more network performance indices through a logical and dynamic process of hierarchy formation [25, 29, 38, 42, 46]. However, the performance of CBR protocols for SC-UWSNs is limited by frequent cluster head (CH) failures and poor cluster formation performance due to the conventional multi-hop data acquisition technique [7, 27, 40, 44].

This paper addresses these cluster-related problems by proposing an energy-efficient hierarchical topology-aware clustering (EEHTAC) protocol for SC-UWSNs. In this paper, fault-tolerant backup clustering (FTBC) algorithms and multi-parameter cluster formation (MPCF) model were developed for the EEHTAC operation. The MPCF model tackles the issue of poor cluster formation performance by integrating multiple parameters to achieve effective clustering process. The FTBC algorithms (FTBC-1 and FTBC-2) tackle the issue of frequent CH failures to avoid interruption in data transmission. FTBC-1 is activated when primary CH failure is discovered and the subsidiary CH is inside coverage of upper layer primary CH, whereas FTBC-2 is triggered when the subsidiary CH is outside coverage, and a bonding CH is elected. Simulations were conducted with OMNET++ 5.4.1 and MATLAB R2018a. Performance of the MPCF model was evaluated using normal, high-fault, and high routing overhead network scenarios. Performance metrics employed for this analysis are temporal topology variation ratio (TTVR), CH load distribution (CLD), and cluster stability (STB). Obtained results show that operating with a CH retention period of 90s achieves better CH duty cycling per round and significantly improves the MPCF process. Performance of the FTBC-based EEHTAC was evaluated relative to Energy-balanced Unequal Layering Clustering (EULC) protocol. Performance indicators adopted for this evaluation are routing overhead ($\Omega$), end to end delay ($\Delta$), CH failures recovered (CFR), CH failures detected (CFD), received packets ($\theta$), and energy consumption ($\Sigma$). With reference to best obtained values, EEHTAC demonstrated significant performance improvement over EULC variants. Obtained results displayed that the MPCF model is highly efficient for cluster formation performance and the FTBC-based EEHTAC protocol can perform effectively well against an existing CBR protocol.

The overall organization of this paper is subsequently given. Review of pertinent related research works is provided in section 2 while section 3 introduces the proposed novel techniques (MPCF, FTBC-1 and FTBC-2) incorporated into the EEHTAC protocol while performance analysis together with simulated results of EEHTAC against variants of EULC CBR protocol is technically discussed in section 4. Section 5 concludes this paper.

## RELATED WORK

The review conducted in this section focuses on CBR protocols implementing depth-based multi-layer hierarchical cluster formation techniques. Clustering-based Geo-opportunistic Depth-based Adjustment Routing (C-GEDAR) was proffered for geo-opportunistic scenarios by relying on cluster formation, sonobuoy-based beaconing, and void node recovery strategies. Evaluation of C-GEDAR's performance exhibited improved performance in terms of coverage recovery and topology control [7, 24]. However, the limitations of this CBR protocol are tied to throughput and energy consumption challenges for UWSN applications demanding periodic monitoring requests.

Shortest Path-based Weighting Depth Forwarding Area Division and Depth Based Routing (SPB-WDFAD-DBR) and Breadth First Shortest Path-based Weighting Depth Forwarding Area Division and Depth Based Routing (BFSPB-WDFAD-DBR) was proposed for maximizing network lifetime by employing breadth-first search (BFS), Dijkstra and coverage radius adjustment algorithms. Simulation results demonstrated better performance with respect to end-to-end delay, packet delivery ratio, and residual energy [13, 30]. High computational costs and packet processing delays are limitations of the proposed CBR protocols.

Software-Defined Clustering Mechanism for Underwater Acoustic Sensor Network (SD-UASN) was recommended as a programmable network infrastructure with SDN switches and controllers by relying on multi-modal cluster communication procedures. Evaluation of SD-UASN's performance showed marked improvement in terms of CH statistics, surviving rate, and coverage ratio [7, 39]. However, SD-UASN suffers from processing overheads as a result of the introduced clustering routines.

AUV-assisted Energy-Efficient Clustering (AEC) was proposed as a network lifespan solution that relies on TDMA clustering, sectoring and AUV traversal strategies. Performance analysis of AEC demonstrated noticeable improvement with respect to currency factor, AUV rounds and residual energy [13, 45]. The major drawback of this protocol is over-dependence on AUV for packet forwarding which leads to lack of flexibility in the mode of data transmission.

Adaptive transmission range in WDFAD-DBR (A-DBR), Cluster-based WDFAD-DBR (C-DBR), Backward transmission-based WDFAD-DBR (B-DBR) and Collision Avoidance-based WDFAD-DBR (CA-DBR) were proffered for optimizing network resources by adopting coverage radius adjustment, backup path selection, collision reduction, and aggregation clustering strategies. Evalauation of these protocols' performance with respect to end-to-end delay, propagation distance, packet delivery ratio, and energy tax exhibited better performance [7, 50]. However, these protocols suffer from long transmission delays and high energy consumption during cluster formation for sparse field of interests (FoIs).

Another technique is the EULC which was proposed as a CBR solution for minimizing and balancing energy consumption in the hot spot region of the UWSN [9]. This was achieved by introducing flexible cluster size formation, periodic collection of sensed data, depth-based unequal layering, and 3D cubic volume deployment of intelligent nodes (self-cognizant of location estimates). Analysis of EULC demonstrated superior performance over DEBCR and LEACH in terms of total packets received, energy tax, CH counts and surviving nodes metrics. However, one of the drawbacks of EULC is that the post-deployment topology does not guarantee optimal coverage. QoS is also not guaranteed as there is limited provisioning for data security and differentiated services.

CBE2R was propounded as a CBR technique for improving link quality, controlling node mobility and efficiently managing battery resources [10]. This technique adopts courier nodes for vertical data transmission, weighted clustering process, and fixed depth-based layering with floatable sink. Performance analysis of CBE2R with respect to residual battery resources, latency, throughput, and network lifespan yielded better performance over DRP, EMGGR, and REEP. However, the limitations of CBE2R are subpar packet delivery ratio (PDR) performance and frequent multiple routing path breakages.

EERBLC was posited as a CBR strategy for minimizing energy dissipation and improving throughput by curbing propagation delays and transmission errors [11]. This strategy employs depth-based unequal layering and clustering, recurrent cluster re-configuration, neighboring sender poll-based routing, autonomous nodes (with homogenous sensor characteristics), and sink connected to continuous source of power supply. Analysis of PDR, network lifespan, duration of stability and throughput performance for EERBLC exhibited superior performance in comparison with EEDBR and DBR. However, the cluster formation process at the network initiation stage of EERBLC introduces high energy tax.

ACUN was recommended as a CBR protocol for curtailing drainage of energy resources as a result of packet transmissions [12]. This protocol implements multi-level hierarchical clustering, hop adjustment routines, routing rules, and intelligent nodes (aware of link state and physicochemical properties of medium). Simulation results of ACUN demonstrated better energy resources utilization, network survivability and network lifespan performance when measured against DEBCR and AFP. However, ACUN suffers from high algorithmic execution costs, increased control packet overheads and critical scalability bottlenecks.

Readers are referred to [7, 13, 24, 30, 39, 41, 43, 45, 50] for a comprehensive review of IoUT/UWSN concepts and state-of-the-art CBR protocols.

## PROPOSED EEHTAC PROTOCOL

This section describes the system model, underlying assumptions, energy consumption model, MPCF model, and FTBC algorithms.

### System model

Figure 1 shows the depth-based multi-layer hierarchical cluster architecture of the proposed EEHTAC routing protocol which comprises of an AUV-based mobile edge entity, dynamic/smart subaquatic sensor nodes and a surface sink. The tapering layered structure introduced in this architecture is to ensure balanced load distribution in the data forwarding chain. The uppermost tier consists of fewer layers than the lower tiers. The uppermost tier is the critical hot spot region. In order to ensure effective load balancing, nodes in the lower tiers are more actively engaged with sensing. The uppermost tier is more actively engaged with forwarding of aggregated data packets to the sink.

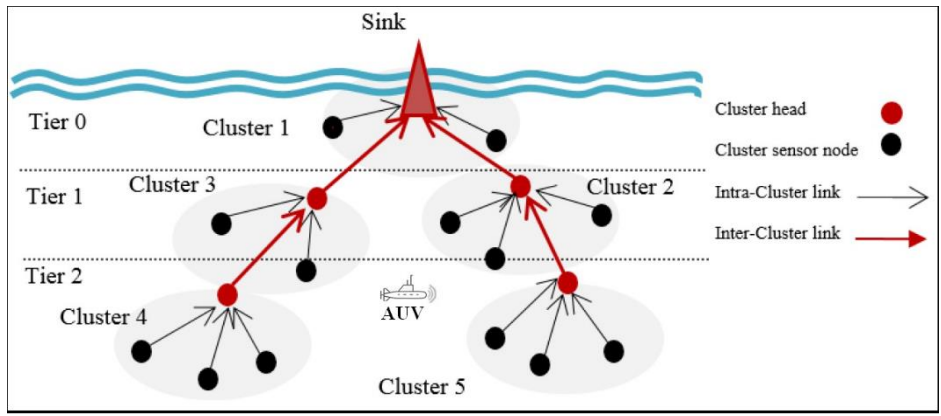

**Figure 1.** SC-UWSN Model for EEHTAC.

### Assumptions

The key assumptions for the proposed EEHTAC protocol are:
1. Initial energy supply for nodes is constant and nodes are resource-constrained.
2. Smart underwater sensor nodes are mobile and affected by water current.
3. Sensed water quality properties (turbidity, pH) are varying.
4. CH performs AUV-to-cluster transmission updates and AUV-assisted TDMA schedule coordination.

The energy consumption model and design requirements adopted in this paper are as described in [9, 17, 18, 21, 22].

### Proposed MPCF model

The MPCF model for the proposed EEHTAC routing protocol relies on aquatic property (AQP), network-level and node-specific parameters. The network-level parameters are:
1. $N$ = total number of smart underwater nodes.
2. $A$ = area of SC-UWSN region.
3. $L_{tot}$ = total layers.

The AQP parameters are:
1. $W_{tbd}$ = turbidity, $0 \leq W_{tbd} \leq 2000$ NTU, admissible range: $W_{tbd} \leq 5$ NTU [14, 15, 28].
2. $W_{PH}$ = pH, $0 \leq W_{PH} \leq 14$, admissible range: $6.5 \leq W_{PH} \leq 9.5$ [14, 16, 28].

The node-specific parameters are:
1. $n$ = smart underwater node entity.

2. $TAG$ = priority tag of n, $1 \leq TAG \leq N$.
3. $E_{init}$ = initial energy of $n_{TAG}$.
4. $E_{rsd}$ = residual energy of $n_{TAG}$, $0 \leq E_{rsd} \leq E_{init}$.
5. $CTR$ = transmission range.
6. $t$ = time period.
7. $H_{nTAG}(t)$ = depth of $n_{TAG}$ at $t$.
8. $T_{TAG}(t)$ = nodes within $n_{TAG}$'s $CTR$, $T_{TAG}(t) \leq N - 1$.
9. $MS_{TAG}(t)$ = mobility of $n_{TAG}$ at $t$.

**Formulation of Electability Factor**

Let $AVB_{TAG}(t)$ be the electability factor of $n_{TAG}$ to be voted as CH at $t$. $A$, $B$, $C$, and $D$ are normalized relative weights of $W_{tbd}$, $W_{PH}$, $E_{rsd}$, $H_{nTAG}$ respectively for controlling the MPCF process performance. $C$, $D$, $ADJ_{TAGk,m}(t)$, and $TCL_{TAG}(t)$ are as derived in Equations (13), (14), (4), and (5) respectively in the Appendix. $AVB_{TAG}$ is derived as:

$$AVB_{TAG}(t) = A \times P_1(t) + B \times P_2(t) + C \times P_3(t) + D \times P_4(t) \tag{1}$$

$$P_1(t) = e^{-MS_{TAG}(t)} \tag{2}$$

$$P_2(t) = \frac{1}{N-1}\sum_{TAG_k=1, TAG_k \neq TAG_m}^{N} ADJ_{TAGk,m}(t) \tag{3}$$

$$P_3(t) = 1 - TCL_{TAG}(t) \tag{4}$$

$$P_4(t) = \begin{cases} \frac{1-TAG}{N}, & Priority = Low\ Tag \\ \frac{TAG}{N}, & Priority = High\ Tag \end{cases} \tag{5}$$

$$A = \frac{|W_{tbd}|norm}{|W_{tbd}|norm+|W_{PH}|norm+|E_{rsd}|norm+|H_{nTAG}|norm} \tag{6}$$

$$B = \frac{|W_{PH}|norm}{|W_{tbd}|norm+|W_{PH}|norm+|E_{rsd}|norm+|H_{nTAG}|norm} \tag{7}$$

$$A + B + C + D = 1;\quad 0 \leq A, B, C, D \leq 1 \tag{8}$$

**Mechanism of the Proposed MPCF Model**

There are four possible scenarios if $n_{TAG}$ is to be voted as CH at $(t + \Delta t)$. These scenarios are:

**Scenario 1:** $n_{TAG}$ assumes CH role if at $t$:
Condition 1: it was not voted earlier as CH ($Q_1(t)$); and
Condition 2: no neighboring CHs are linked to it ($Q_2(t)$); and
Condition 3: its $AVB_{TAG}$ is $> AVB_{TAG}$ of all uncovered lower priority tagged neighbors ($Q_3(t)$); and Condition 4: its $AVB_{TAG}$ is $\geq AVB_{TAG}$ of all uncovered higher priority tagged neighbors ($Q_4(t)$)

**Scenario 2:** $n_{TAG}$ keeps being a CH if at $t$:
Condition 1: it was earlier voted as CH ($Q_5(t)$); and
Condition 2: its $AVB_{TAG}$ is $> AVB_{TAG}$ of all lower priority tagged neighboring CHs ($Q_6(t)$); and
Condition 3: its $AVB_{TAG}$ is $\geq AVB_{TAG}$ of all higher priority tagged neighboring CHs ($Q_7(t)$)

**Scenario 3:** $n_{TAG}$ replaces a neighboring CH if at $t$:
Condition 1: it was not voted earlier as CH ($Q_8(t)$); and
Condition 2: it has neighboring CH that is $\geq 1$ ($Q_9(t)$); and
Condition 3: subtraction of its $AVB_{TAG}$ from $AVB_{TAG}$ of all lower priority tagged neighboring CHs is $> AVB_{SET}$ ($Q_{10}(t)$); and Condition 4: subtraction of its $AVB_{TAG}$ from $AVB_{TAG}$ of all higher priority tagged neighboring CHs is $\geq AVB_{SET}$ ($Q_{11}(t)$)

**Scenario 4:** $n_{TAG}$ takes up CM role in all other scenarios.

From Scenario 3, Conditions 3 and 4, $AVB_{SET}$ means the set electability threshold. This abovementioned MPCF process is formulated through the four scenarios as:

$$CH_{TAG}(t+\Delta t) = Q_1(t)Q_2(t)Q_3(t)Q_4(t) + Q_5(t)Q_6(t)Q_7(t) + Q_8(t)Q_9(t)Q_{10}(t)Q_{11}(t) \tag{9}$$

$$Q_1(t) = Q_8(t) = 1 - CH_{TAG_m}(t) \tag{10}$$

$$Q_2(t) = 1 - U\left[\sum_{TAG_k=1, TAG_k \neq TAG_m}^{N}\left(CH_{TAG_k}(t) \times ADJ_{TAGk,m}(t)\right)\right] \tag{11}$$

$$Q_3(t) = \prod_{TAG_k=1}^{TAG_m-1} U\left[AVB_{TAG_m}(t) - ADJ_{TAGk,m}(t)AVB_{TAG_k}(t)(1 - COV_{TAG_k}(t))\right] \tag{12}$$

$$Q_4(t) = \prod_{TAG_k=TAG_m+1}^{N} 1 - U\left[ADJ_{TAGk,m}(t)AVB_{TAG_k}(t)(1 - COV_{TAG_k}(t)) - AVB_{TAG_m}(t)\right] \tag{13}$$

$$Q_5(t) = CH_{TAG_m}(t) \tag{14}$$

$$Q_6(t) = \prod_{TAG_k=1}^{TAG_m-1} U[AVB_{TAG_m}(t) - ADJ_{TAG_{k,m}}(t)AVB_{TAG_k}(t)CH_{TAG_k}(t)] \quad (15)$$

$$Q_7(t) = \prod_{TAG_k=TAG_m+1}^{N} 1 - U[ADJ_{TAG_{k,m}}(t)AVB_{TAG_k}(t)CH_{TAG_k}(t) - AVB_{TAG_m}(t)] \quad (16)$$

$$Q_9(t) = U\left[\sum_{TAG_k=1,TAG_k \neq TAG_m}^{N} CH_{TAG_k}(t)ADJ_{TAG_{k,m}}(t)\right] \quad (17)$$

$$Q_{10}(t) = \prod_{TAG_k=1}^{TAG_m-1} U[AVB_{TAG_m}(t) - ADJ_{TAG_{k,m}}(t)(AVB_{TAG_k}(t) + AVB_{SET})CH_{TAG_k}(t)] \quad (18)$$

$$Q_{11}(t) = \prod_{TAG_k=TAG_m+1}^{N} 1 - U[ADJ_{TAG_{k,m}}(t)(AVB_{TAG_k}(t) + AVB_{SET})CH_{TAG_k}(t) - AVB_{TAG_m}(t)] \quad (19)$$

$$U[x] = \begin{cases} 1, & x > 0 \\ 0, & x \leq 0 \end{cases} \quad (20)$$

Table 1 describes the algorithm for the MPCF model.

**Table 1.** Algorithm for the MPCF Model.

```
initialize AVBSET, N
  for each ordinary node n_TAG
      compute Q₁, Q₂, Q₃, Q₄
  /*If Scenario 1 holds, Q₁, Q₂, Q₃, Q₄ fires with value → 1 while
    others → 0*/
      compute Q₅, Q₆, Q₇
  /*If Scenario 2 holds, Q₅, Q₆, Q₇ fires with value → 1 while others
    0*/
      Q₈ ← Q₁
      compute Q₉, Q₁₀, Q₁₁
  /*If Scenario 3 holds, Q₈, Q₉, Q₁₀, Q₁₁ fires with value → 1  while
    others → 0*/
  end for
    /*The Scenarios 1, 2, 3 and 4 are independent and separated with
      OR (+) operation while the dependent conditions under each
      scenario are related with AND (*) operation. The new CH
      electability is thus computed.*/
      new CH_TAG = (Q₁*Q₂*Q₃*Q₄)+(Q₅*Q₆*Q₇)+(Q₈*Q₉*Q₁₀*Q₁₁)
    /*In Scenario 4 where all Q-Parameters fails to fire, the node
      assumes cluster member status*/
```

## Cluster setup phase

This phase begins with advertisement process for an allotted period $t_{ADV}$. Advertisement process commences with each node broadcasting HELLO packet. The HELLO packet contains the electability factor, residual energy, priority tag and layer number. Upon reception of HELLO packets, each node retains packets coming from neighbors on the same layer while discarding packets coming from nodes in other layers.

After $t_{ADV}$ has elapsed, CH contention process is automatically activated. Therefore, the MPCF function is called for nodes whose initial $\Delta t$ (retention period) have elapsed in order to compute their respective $AVB_{TAG}$ (electability factor). Primary and Subsidiary CHs are thereafter voted and chosen. Subsequently, POLLING packets are broadcasted by the newly-voted Primary CHs. POLLING packet includes $AVB_{TAG}$, competition cluster radius, layer number, priority tag, and residual energy information.

Non-CH nodes respond to POLLING packets with JOIN packets to their respective nearest Primary CHs. JOIN packet contains residual energy of non-CH, priority tag of nearest Primary CH, and priority tag of non-CH. Primary CHs enlist their respective non-CH as CMs upon reception of JOIN packets. Afterwards, data transmission and AUV-assisted data collection is performed before this next round of network operation. The cluster setup algorithm for the proposed EEHTAC routing protocol is described in Table 2.

Table 2. Cluster Setup Algorithm.

```
for each node n_TAG
    compute layer number L_nTAG
    n_TAG ← L_nTAG
end for
for each node n_TAG
    broadcast HELLO
end for
for each node n_TAG received HELLO packet
    if node's L_nTAG in received packet = n_TAG's L_nTAG
        keep INFO of neighbors
        else drop packet
    end if
end for

for each node n_TAG
    if  L_nTAG ≠ 1
        n_TAG compute Δt
    end if
end for
for each node n_TAG
            while Δt does not expire
               end while
    if did not receive  POLLING packet
        call function MPCF
        elect(Primary CH) ← ACTIVE
        elect(Subsidiary CH) ← IDLE
        store electability factor AVB_TAG
        compute competition cluster radius Rad_cmp
        broadcast POLLING packet within Rad_cmp
    else update(Primary CH) ← POLLING packet
  end if
end for
for each ordinary node received POLLING packet
    compute relative distance cost
    send JOIN packet
end for
for each CH
    register CM list
end for
```

However, this cluster setup phase is however often interrupted and disrupted by frequent CH failures. The yardstick for failure detection is if $E_{rsd}$ of CH is $\leq E_{surv}$; or if number of retransmitted JOIN packets $\geq SRL$. $E_{surv}$ is the survivability threshold and *SRL* is the set retransmission limit. In order to ensure effective detection and recovery from CH failures, FTBC-1 and FTBC-2 algorithms are introduced in this paper.

**FTBC-1 Algorithm**

FTBC-1 algorithm is triggered upon detection of non-reception of ACK response from Primary CH till the *time_to_join* expires which is a reliable indication of Primary CH failure. Subsequently, the Subsidiary CH is put to wake state for the purpose of each ordinary node without ACK response (provided the counts of retransmitted JOIN packet has exceeded the set retransmission limit). The status of the Subsidiary CH is also probed to check if the Subsidiary CH is still within transmission radius of upper layer Primary CH and its residual energy is greater than the survivabilty threshold. After successful checking, the status of the Subsidiary CH is updated as Primary CH and POLLING packets are broadcasted within the competition cluster radius of the new Primary CH. Upon reception of the POLLING packets, the ordinary nodes re-send JOIN packets to the new Primary CH to be registered as CM by the new Primary CH. Table 3 provides a procedural description of the FTBC-1 algorithm.

Table 3. FTBC-1 Algorithm.

```
Step 1 – 36 of function cluster_configuration
NO ACK response from Primary CH till time_to_join expires
   for each ordinary node with NO ACK reply
      if (counts of retransmitted JOIN packet > retransmission_limit) || (E_rsd ≤ E_surv)
          wake Subsidiary CH
          probe Subsidiary CH
        if (E_rsd of Subsidiary CH ≥ E_surv) && (Rad_cmp links Primary CH in L_nTAG - 1)
            update status: Primary CH ← Subsidiary CH
            broadcast POLLING packet within Rad_cmp
        end if
         else call function FTBC_TWO
      end if
      re-send JOIN packet to new Primary CH
   end for
       for each new Primary CH
           register updated CM list
         end for
```

## FTBC-2 Algorithm

FTBC-2 algorithm is triggered when Subsidiary CHs falls out of transmission radius of upper layer Primary CHs. In this case, Bonding CH is voted and chosen by Subsidiary CH based on the MPCF process. The status of the Bonding CH is probed, its electability factor is stored and its competition cluster radius is computed to check if the Bonding CH is within transmission radius of upper layer Primary CH and its residual energy is greater than the survivabilty threshold. Subsequently, the Bonding CH is activated for the purpose of each ordinary node without ACK response from the Subsidiary CH (provided the counts of retransmitted JOIN packet has exceeded the set retransmission limit). After successful checking, the status of the Bonding CH is updated as Primary CH and POLLING packets are broadcasted within its competition cluster radius of the new Primary CH. Upon reception of the POLLING packets, the ordinary nodes re-send JOIN packets to the new Primary CH to be registered as CM by the new Primary CH. Table 4 gives the step-by-step description of FTBC-2 algorithm.

Table 4. FTBC-2 Algorithm.

```
Step 1 – 36 of function cluster_configuration
Step 3 – 7 of function FTBC_ONE
    if ~(E_rsd of Subsidiary CH ≥ E_surv) || ~(Rad_cmp links Primary CH in L_nTAG - 1)
        call function MPCF
        elect(Bonding CH) ← ACTIVE
        store electability factor AVB_TAG
        compute competition cluster radius Rad_cmp
        Probe Bonding CH
   for each ordinary node with NO ACK reply from Subsidiary CH
       if (E_rsd of Bonding CH ≥ E_surv) && (Rad_cmp links Primary CH in L_nTAG - 1)
            update status: Primary CH ← Bonding CH
            broadcast POLLING packet within Rad_cmp
         else repeat Step 5 - 9
       end if
        re-send JOIN packet to new Primary CH
    end for
     end if
        for each new Primary CH
            register updated CM list
          end for
```

## Application Area of Proposed EEHTAC Protocol

The application area of relevance of the proposed EEHTAC protocol is in freshwater quality monitoring and surveillance (FWQMS). This proposed protocol is specifically designed for FWQMS applications for determining the existence of contaminants that are harmful to aquatic habitats.

## RESULTS AND DISCUSSIONS

This section discusses results obtained from formulating the MPCF model and incorporating the FTBC-1 and FTBC-2 algorithms into the operation of the proposed EEHTAC routing protocol.

## Performance metrics

The measures employed for performance analysis are cluster stability (STB), energy consumption (Σ), received packets (θ), temporal topology variation ratio (TTVR), CH failures detected (CFD), CH load distribution (CLD), and CH failures recovered (CFR), routing overhead (Ω), and end-to-end delay (Δ).

## Metrics Definition

1. Σ-metric measures the aggregate energy consumption for the smart underwater nodes after a stipulated network round ($R$).
2. θ-metric determines the total received packets at surface sink from the uppermost-tier nodes after $R$ operations.
3. CFD-metric quantifies the ratio (in %) of detected CH failures at $R$ operations.
4. CFR-metric estimates the ratio (in %) of recovered CH failures at $R$ operations.
5. STB-metric determines the frequency of enlisting and delisting CHs for the smart underwater nodes after $R$ operations.
6. CLD-metric measures the load sharing burden of CHs (AUV-to-cluster data forwarding chain inclusive) after $R$ operations.
7. TTVR-metric is the ratio of mean node drifts to the modal value of node drifts over $R$ operations.
8. Ω-metric is the ratio of packet processing time taken prior to the successful execution of routing commands to the network operation time after $R$ operations.
9. Δ-metric measures the aggregate delay for node-to-sink data transmissions after $R$ operations.

## Simulation parameters

Simulations were conducted with OMNET++ 5.4.1 and MATLAB R2018a. The parameters utilized in this paper for the simulation experiment are 500 deployed nodes, 500m x 500m x 500m network size, sink location at surface center, average packet size of 500 bytes, data rate of 4 kbps, initial energy of 2J, idle energy of 0.1µJ, fusing energy of 5nJ/bit, AUV speed of 2.0 m/s, carrier frequency of 27 kHz, and critical transmission range of 60m.

## Analysis of MPCF performance

High-fault network condition ($\Delta t = 0.1$s), normal network condition ($\Delta t = 90$s), and high routing overhead condition ($\Delta t = 300$s) are the three network scenarios used to analyze the performance of the MPCF model.

### Effect of CH Retention Period (Δt) on Cluster Stability (STB)

From Table 5, a higher *STB* value ($STB \geq 0.45$) means the CH assignment is switched for the smart underwater nodes at a desirably low frequency. In the high-fault network condition, the network recorded cluster stability for 30% of operational period as it was only able to reach the *STB* threshold after 42 minutes out of 60 minutes of network functioning. In the normal network condition, the network displayed cluster stability for 60% of operational period as it was able to meet the threshold after 24 minutes of running the network. The high routing overhead condition exhibits similar trend but the high *STB* values under this condition are as a result of the long delay in re-enlisting and switching CHs due to congestion and latency issues. The results presented in Table 5 underscore the importance of adopting proper CH duty cycling on the quality of cluster stability performance.

Table 5. Effect of CH Retention Period (Δt) on Cluster Stability (STB).

|       | $\Delta t$ |        |        |
|-------|--------|--------|--------|
| $R$   | 0.1    | 90     | 300    |
| 0     | 0.0271 | 0.0772 | 0.0913 |
| 100   | 0.1035 | 0.1282 | 0.1466 |
| 200   | 0.1708 | 0.2019 | 0.2574 |
| 300   | 0.3062 | 0.4508 | 0.4893 |
| 400   | 0.3747 | 0.6201 | 0.6619 |
| 500   | 0.4219 | 0.7422 | 0.7936 |
| 600   | 0.4614 | 0.6175 | 0.6884 |
| 700   | 0.5005 | 0.5823 | 0.5986 |
| 800   | 0.4301 | 0.4418 | 0.4569 |

### Effect of CH Retention Period (Δt) on CH Load Distribution (CLD)

From Table 6, a higher *CLD* value ($CLD \geq 0.50$) is desired as it indicates load sharing for CHs (AUV-to-cluster data forwarding chain inclusive) is higher over $R$ operations. In the high-fault network condition, the network achieved CH load balancing for 10% of operational period as it was only able to briefly reach the *CLD* threshold from 36th minute till 42nd minute. In the normal network condition, the network maintained CH load balancing for 55.56% of operational period before maintaining a stable *CLD* state as it was able to operate within the threshold from $R = 300$ to $R = 700$. In the high routing overhead condition, the network was not able to reach the threshold. The results presented in Table 6 show the effectiveness of selecting proper CH duty cycling on the quality of CH load distribution performance.

Table 6. Effect of CH Retention Period ($\Delta t$) on CH Load Distribution (CLD).

|   | $\Delta t$ | | |
|---|---|---|---|
| $R$ | 0.1 | 90 | 300 |
| 0 | 0.0415 | 0.0837 | 0.0326 |
| 100 | 0.1408 | 0.2981 | 0.1054 |
| 200 | 0.1936 | 0.3729 | 0.1457 |
| 300 | 0.2523 | 0.5814 | 0.2031 |
| 400 | 0.3915 | 0.5627 | 0.2886 |
| 500 | 0.5143 | 0.6988 | 0.3706 |
| 600 | 0.3776 | 0.6152 | 0.2901 |
| 700 | 0.2403 | 0.5309 | 0.1805 |
| 800 | 0.1682 | 0.3489 | 0.1086 |

*Effect of CH Retention Period (Δt) on Temporal Topology Variation Ratio (TTVR)*

From Table 7, a lower *TTVR* value ($0.25 \leq TTVR \leq 0.40$) is desired as it is an indication of suitably smaller drifts out of the competition cluster radius during network operation after $R$ rounds. In the high-fault network condition, the network displayed topology stability for 22.22% of operation period as it was able to function limitedly in the *TTVR* threshold from 12th minute till 18th minute. In the normal network condition, the network recorded topology stability for 66.67% of operation period before reaching a stable *TTVR* state as it was able to function longer within the threshold from R = 300 to R = 800. In the high routing overhead condition, the network was not able to reach the threshold. The results presented in Table 7 demonstrate the importance of employing proper duty cycling on the quality of topology stability performance.

Table 7. Effect of CH Retention Period ($\Delta t$) on Temporal Topology Variation Ratio (TTVR).

|   | $\Delta t$ | | |
|---|---|---|---|
| $R$ | 0.1 | 90 | 300 |
| 0 | 0.1583 | 0.0642 | 0.0107 |
| 100 | 0.2531 | 0.1689 | 0.0866 |
| 200 | 0.3879 | 0.2176 | 0.1074 |
| 300 | 0.5066 | 0.2834 | 0.1491 |
| 400 | 0.6747 | 0.3262 | 0.2109 |
| 500 | 0.7839 | 0.3952 | 0.1683 |
| 600 | 0.6433 | 0.3417 | 0.1104 |
| 700 | 0.5428 | 0.2986 | 0.0968 |
| 800 | 0.4521 | 0.2503 | 0.0319 |

The summary of MPCF performance evaluation results is given in Table 8.

Table 8. MPCF Performance Evaluation Summary.

| Metric | Desired Threshold | $\Delta t$ | | |
|---|---|---|---|---|
|  |  | 0.1 | 90 | 300 |
| STB | $\geq 0.45$ | 30% | 60% | 60% |
| CLD | $\geq 0.50$ | 10% | 55.56% | 0% |
| TTVR | 0.25 - 0.40 | 22.22% | 66.67% | 0% |

**Comparative performance evaluation of EEHTAC (based on FTBC-1 & FTBC-2)**

The EULC technique was selected as baseline protocol in this paper. Four EULC variants were used with respect to the weights for cluster configuration as described in [9] as shown in Table 9.

Table 9. EULC Variants.

| Variant | α | β | γ |
|---------|-----|-----|-----|
| EULC-1 | 0.2 | 0.3 | 0.5 |
| EULC-2 | 0.1 | 0.5 | 0.4 |
| EULC-3 | 0.4 | 0.3 | 0.3 |
| EULC-4 | 0.4 | 0.4 | 0.2 |

*Evaluation of Energy Consumption (Σ) Performance*

Figure 2 provides the comparative plot of $\Sigma$ for EEHTAC against EULC variants. It was observed that EEHTAC has lesser energy tax by yielding an improvement of 25.84%, 45.60%, 54.35%, and 13.98% over EULC-1, EULC-2, EULC-3, and EULC-4, respectively. MPCF cluster maintenance benefits together with FTBC-1 and FTBC-2's resource-aware adaptive clustering benefits justify the reduction in EEHTAC's energy tax. Another justification for the improved energy consumption performance of EEHTAC is the stable and flexible topology control for cluster-based communications which signifcantly reduces execution time for the deployment scenario adopted in this UWSN application.

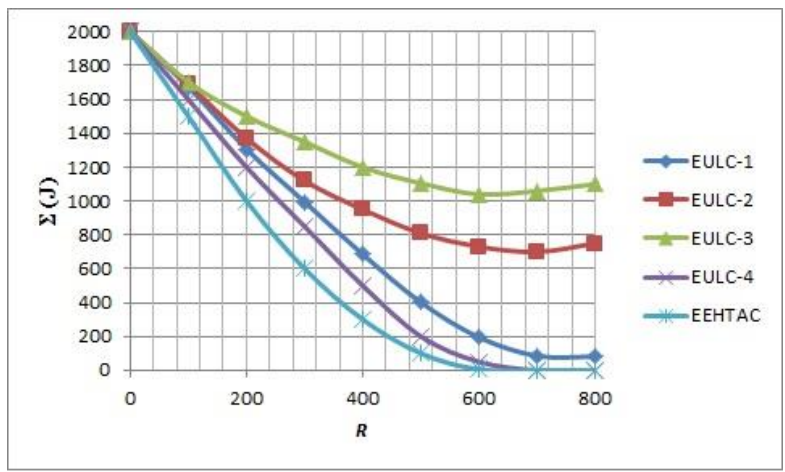

**Figure 2.** Comparative Plot of Energy Consumption ($\Sigma$).

*Evaluation of Received Packets (θ) Performance*

Figure 3 provides the comparative plot of $\theta$ for EEHTAC against EULC variants. It was observed that EEHTAC displayed more packet reception by yielding an improvement of 67.34%, 16.13%, 86.65%, and 51.27% over EULC-1, EULC-2, EULC-3, and EULC-4, respectively. Adoption of AUV-assisted data collection, cluster stabilizing routines and adaptive ternary CH (Bonding, Subsidiary and Primary CHs) backup process justifies the high packet reception at sink for the EEHTAC protocol. Additionally, the robust and fault-tolerant cluster management process improves network throughput and packet delivery which is another justification for the enhanced received packets performance.

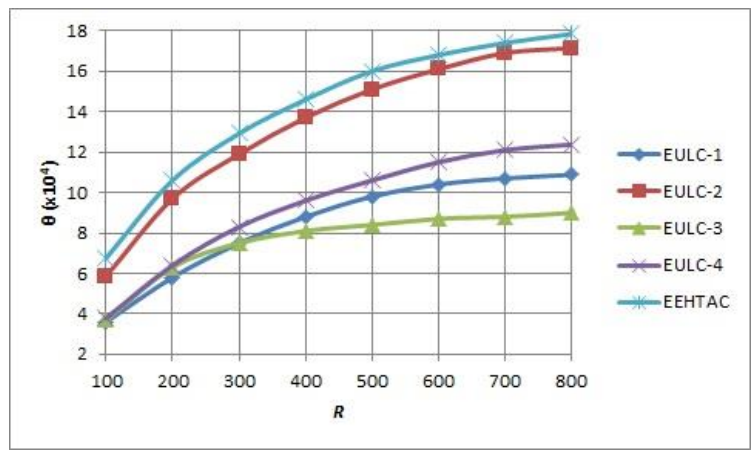

**Figure 3.** Comparative Plot of Received Packets ($\theta$).

*Evaluation of CH Failures Detected (CFD) Performance*

Figure 4 provides the comparative plot of *CFD* for EEHTAC against EULC variants. It was observed that EEHTAC recorded lesser cases of CH failures by exhibiting an improvement of 28.02%, 24.73%, 17.40%, and 10.09% over EULC-1, EULC-2, EULC-3, and EULC-4, respectively. Justifications for this improvement are the CH backup and recovery mechanisms of FTBC-1 and FTBC-2. Further justifications for the improved CFD performance is the adaptive and flexible CH allocation and backup process which makes it possible to achieve seamless CH-based data transmissions which signifcantly reduces cases of detected CH failures for the deployment scenario adopted in this UWSN application.

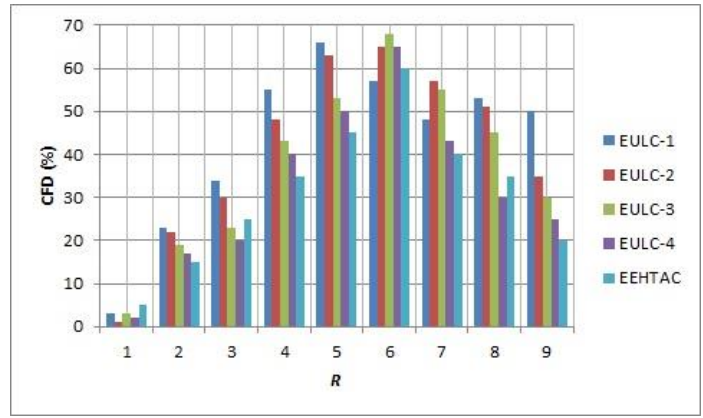

**Figure 4.** Comparative Plot of CH Failures Detected (*CFD*).

*Evaluation of CH Failures Recovered (CFR) Performance*

Figure 5 provides the comparative plot of *CFR* for EEHTAC against EULC variants. It was observed that EEHTAC recorded more cases of recovery from CH failures by yielding an improvement of 81.33%, 79.12%, 71.08%, and 66.57% over EULC-1, EULC-2, EULC-3, and EULC-4, respectively. Justifications for this improvement are the algorithmic enhancements of FTBC-1 and FTBC-2. Further justifications for the improved CFR performance is the fault-tolerant CH status monitoring and recovery process which makes it quicker to attain CH-based data transmissions with minimal failures leading to signifcant reduction in cases of recovered CH failures for the deployment scenario adopted in this UWSN application.

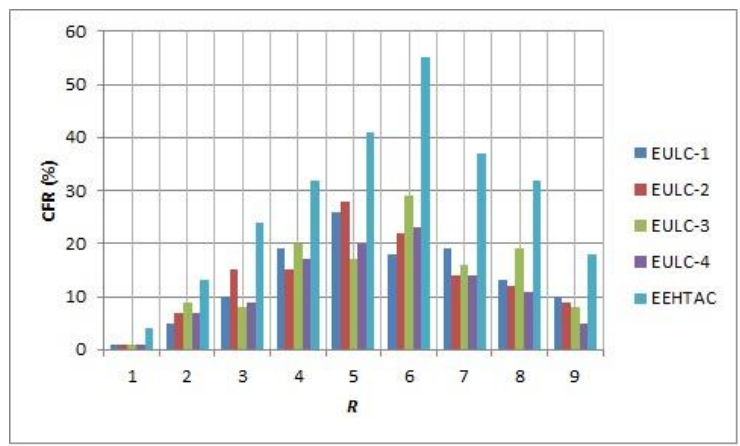

**Figure 5.** Comparative Plot of CH Failures Recovered (*CFR*).

*Evaluation of End-To-End Delay (Δ) Performance*

Figure 6 provides the comparative plot of *Δ* for EEHTAC against EULC variants. It was observed that EEHTAC recorded marked improvements in the end-to-end delay results by yielding an improvement of 16.94%, 11.06%, 29.94%, and 8.60% over EULC-1, EULC-2, EULC-3, and EULC-4, respectively. Justifications for this improvement are the algorithmic enhancements of FTBC-1 and FTBC-2. Additional justifications for the improved end-to-end delay performance is the flexible and adaptive maintenance of CM-to-CH, CH-to-CH and CH-to-Sink data forwarding chains which makes it easier to repair and preserve end-to-end communication links within the UWSN.

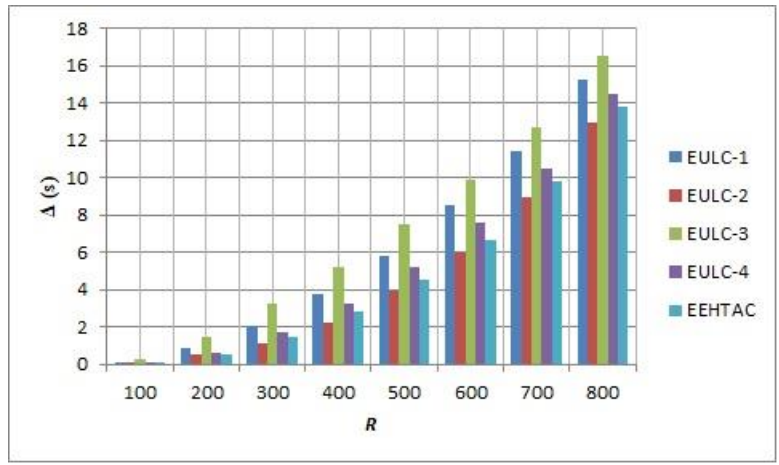

**Figure 6.** Comparative Plot of End-To-End Delay ($\Delta$).

*Evaluation of Routing Overhead ($\Omega$) Performance*

Figure 7 provides the comparative plot of $\Omega$ for EEHTAC against EULC variants. It was observed that EEHTAC recorded less cases of routing overhead by yielding an improvement of 51.71%, 41.77%, 18.85%, and 58.40% over EULC-1, EULC-2, EULC-3, and EULC-4, respectively. Justifications for this improvement are the algorithmic enhancements of FTBC-1 and FTBC-2. Further justifications for the improved routing overhead performance is the adaptive and flexible CH allocation and backup process which makes it possible to achieve seamless CH-based data transmissions with signifcantly reduced cases of routing overhead.

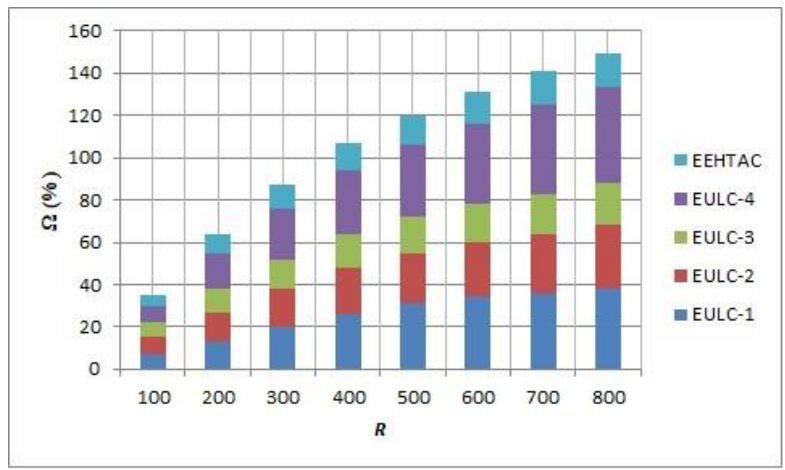

**Figure 7.** Comparative Plot of Routing Overhead ($\Omega$).

Table 10 summarizes the comparative performance evaluation of EEHTAC against EULC variants.

**Table 10.** Summary of EEHTAC Performance Improvements over EULC Variants.

| Metric | EULC-1 | EULC-2 | EULC-3 | EULC-4 |
| --- | --- | --- | --- | --- |
| $\Sigma$ | 25.84% | 45.60% | 54.35% | 13.98% |
| $\theta$ | 67.34% | 16.13% | 86.65% | 51.27% |
| CFD | 28.02% | 24.73% | 17.40% | 10.09% |
| CFR | 81.33% | 79.12% | 71.08% | 66.57% |
| $\Delta$ | 16.94% | 11.06% | 29.94% | 8.60% |
| $\Omega$ | 51.71% | 41.77% | 18.85% | 58.40% |

**CONCLUSION**

This paper tackles the issues of poor cluster formation performance and frequent cluster head failures for SC-UWSN by developing an energy-efficient hierarchical topology-aware clustering routing (EEHTAC) protocol. In this paper, fault-tolerant backup clustering (FTBC) algorithms and multi-parameter cluster formation (MPCF) model were developed for the EEHTAC operation. The MPCF model tackles the issue of poor cluster formation performance by integrating multiple parameters to achieve effective clustering process. The FTBC algorithms (FTBC-1 and FTBC-2) tackle the issue of frequent CH failures to avoid interruption in data transmission. FTBC-1 is activated when primary CH failure is discovered and the subsidiary CH is inside coverage of upper layer primary CH, whereas FTBC-2 is triggered when the subsidiary CH is outside coverage, and a bonding CH is elected. Simulations were conducted with OMNET++ 5.4.1 and MATLAB R2018a. Performance of the MPCF model was evaluated using normal, high-fault, and high routing overhead network scenarios. Performance metrics employed for this analysis are temporal topology variation ratio (TTVR), CH load distribution (CLD), and cluster stability (STB). Obtained results show that operating with a CH retention period of 90s achieves better CH duty cycling per round and significantly improves the MPCF process. Performance of the FTBC-based EEHTAC was evaluated relative to Energy-balanced Unequal Layering Clustering (EULC) protocol. Performance indicators adopted for this evaluation are routing overhead ($\Omega$), end to end delay ($\Delta$), CH failures recovered (CFR), CH failures detected (CFD), received packets ($\theta$), and energy consumption ($\Sigma$). With reference to best obtained values, EEHTAC demonstrated significant performance improvement over EULC variants. Obtained results displayed that the MPCF model is highly efficient for cluster formation performance and the FTBC-based EEHTAC protocol can perform effectively well against an existing CBR protocol. Future work will develop a dynamic link repair scheme for efficient detection and recovery from cluster coverage issues and an energy-efficient cluster management technique for reducing cluster management overhead.

## ACKNOWLEDGMENTS

The authors would like to thank Ahmadu Bello University, Zaria for their support and affording the resources to complete this research work. The authors would also like to thank the journal's editorial team of reviewers for their constructive comments which has helped to improve the technical quality of this research work.

# APPENDIX

## Hierarchical Clustering Parameters Derivation:

Let $\Delta t$ represent the CH retention period which is contrived as:

$$\Delta t = t_{cmp} \times \left[\frac{E_{rsd}}{E_{init}} - 1\right] + [rnd \times t_{cmp}] \tag{1}$$

From Equation (1), $t_{cmp}$ is the allotted period to operate as CH and $rnd$ is a random value (0.5 – 1).

Let $CH_{TAG}(t)$ denote that $n_{TAG}$ is a CH at $t$, which is expressed as:

$$CH_{TAG}(t) = \begin{cases} 1, & if\ n_{TAG} = CH\ at\ t \\ 0, & if\ n_{TAG} \neq CH\ at\ t \end{cases} \tag{2}$$

Let $COV_{TAG}(t)$ indicate that $n_{TAG}$ is covered by a CH at $t$, which is expressed as:

$$COV_{TAG}(t) = \begin{cases} 1, & if\ n_{TAG} \in CH(>1)\ at\ t \\ 0, & if\ n_{TAG} \notin CH(=0)\ at\ t \end{cases} \tag{3}$$

Let $ADJ_{TAGk,m}(t)$ indicate that $n_{TAGk}$ is adjacent to $n_{TAGm}$ at $t$. This relationship is represented as:

$$ADJ_{TAGk,m}(t) = \begin{cases} 1, & if\ n_{TAGk} \cup n_{TAGm}\ at\ t \\ 0, & if\ n_{TAGk} \cup n_{TAGm}\ at\ t \end{cases} \tag{4}$$

Let $TCL_{TAG}(t)$ be the fraction of time $n_{TAG}$ remains as CH for a relatively long period ($T$) of network operation. This parameter is formulated as:

$$TCL_{TAG}(t) = \frac{1}{k}\sum_{a=1}^{k} CH_{TAG}(t - a\Delta t);\quad 0 \leq TCL_{TAG}(t) \leq 1\ \&\ T = k \times \Delta t \tag{5}$$

Let $STB_{TAG}(t)$ represent the stability measure for the number of times $n_{TAG}$ is elected and remains as CH for $T$ period. This stability parameter is formulated as:

$$STB_{TAG}(t) = e^{-L_{TAG}(t)};\quad 0 \leq STB_{TAG}(t) \leq 1 \tag{6}$$

Where;

$$L_{TAG}(t) = \frac{1}{k\Delta t}\sum_{a=1}^{k}|CH_{TAG}(t-(a-1)\Delta t) - CH_{TAG}(t-a\Delta t)|;\quad T = k \times \Delta t \tag{7}$$

Let $DCH(t)$ be the distribution of CHs in the SC-UWSN region which is derived as:

$$DCH(t) = \frac{1}{N}\sum_{TAG=1}^{N} TCL_{TAG}(t);\quad 0 \leq DCH(t) \leq 1 \tag{8}$$

Let $STB(t)$ be the cluster stability which is contrived as:

$$STB(t) = \frac{1}{N}\sum_{TAG=1}^{N} STB_{TAG}(t);\quad 0 \leq STB(t) \leq 1 \tag{9}$$

Let $CLD(t)$ be the CH load distribution for the SC-UWSN which is formulated as:

$$CLD(t) = 1 - \sqrt{\frac{1}{N}\sum_{TAG}^{N}(TCL_{TAG}(t) - DCH(t))^2};\quad 0 \leq CLD(t) \leq 1 \tag{10}$$

## Depth-based Layering Parameters Derivation

Let $L_{nTAG}(t)$ represent the layer number of $n_{TAG}$ at $t$ which is derived as:

$$L_{nTAG}(t) = \left[\frac{H_{nTAG}}{CTR}\right] + k_{layer} \tag{11}$$

From Equation (11), $k_{layer}$ is the layering constant ($1 \leq k_{layer} \leq 5$). Let $Rad_{cmp}(t)$ represent the competition cluster radius. This is formulated as:

$$Rad_{cmp}(t) = \left[C \times \frac{E_{rsd}}{E_{init}}\right] \times \left[CTR \times D \times \frac{L_{nTAG}(t)}{L_{tot}}\right] \tag{12}$$

$$C = \frac{|E_{rsd}|norm}{|W_{tbd}|norm + |W_{PH}|norm + |E_{rsd}|norm + |H_{nTAG}|norm} \tag{13}$$

$$D = \frac{|H_{nTAG}|norm}{|W_{tbd}|norm + |W_{PH}|norm + |E_{rsd}|norm + |H_{nTAG}|norm} \tag{14}$$

$Rad_{cmp}$ considers $C$ and $D$ which are normalized relative weights of $E_{rsd}$, and $H_{nTAG}$ respectively.